\documentclass[12pt,preprint]{aastex}

                   \newcommand{\ths}{\thinspace}
                   
                   \newcommand{\tle}{{\raise 0.3ex\hbox{$\sc {\ths \le \ths }$}}}
                   \newcommand{\tge}{{\raise 0.3ex\hbox{$\sc {\ths \ge \ths }$}}}
                   \newcommand{\tl}{{\raise 0.3ex\hbox{$\sc {\ths < \ths }$}}}
                   \newcommand{\tg}{{\raise 0.3ex\hbox{$\sc {\ths > \ths }$}}}
                   \newcommand{\ts}{{\raise 0.3ex\hbox{$\sc {\ths \sim \ths }$}}}
                   \newcommand{\Msun}{\hbox{$\ths M_{\odot}$}}
                   \shorttitle{Global dynamics of planetary systems}
                   \shortauthors{Kiseleva-Eggleton et al.}

\begin{document}

                   \title{Global dynamics and stability limits for planetary systems around
                    HD$\thinspace$12661, HD$\thinspace$38529,  HD$\thinspace$37124 and HD$\thinspace$160691.}
                   \author{Ludmila Kiseleva-Eggleton\altaffilmark{1,2}} 
                   \affil{Department of Mathematics and Computer Science, School of Science, St.
                    Mary's College of California, Moraga, CA 94575, USA}
                   \email{lkissele@stmarys-ca.edu}
                   \author{ Eric Bois \& Nicolas Rambaux}
                   \affil{Observatoire de Bordeaux, 2 rue d'Observatoire, B.P.89, F-33270,
                   Floirac, France}
                   \email{bois@observ.u-bordeaux.fr; rambaux@observ.u-bordeaux.fr}
                   \and
                   \author{Rudolf Dvorak}
                   \affil{Institute of Astronomy, University of Vienna, T\"{u}rkenschanzstrasse 17,
                    A-1180 Vienna, Austria}
                   \email{dvorak@astro.univie.ac.at}
		   \altaffiltext{1}{This work was carried out at Observatoire de Bordeaux
		   (Universit\'e de Bordeaux I), France}
                   \altaffiltext{2}{Also a participating guest at IGPP, LLNL, L-413, 7000 East Ave,
                    Livermore, CA 94550, USA}

                   \begin{abstract}
                   In order to distinguish between regular and chaotic planetary
                    orbits we apply a new 
                   technique called MEGNO in a wide neighbourhood of orbital parameters determined using 
                   standard two-body Keplerian fits for HD$\thinspace$12661, HD$\thinspace$38529, HD$\thinspace$37124,
                   HD$\thinspace$160691
                    planetary systems. We 
                   show that the currently announced orbital parameters place these systems in very
                    different situations 
                   from the point of view of dynamical stability. While HD$\thinspace$38529 and HD$\thinspace$37124 are
                    located within large stability
                    zones in the phase space around their determined orbits, the preliminary orbits
                    in  HD$\thinspace$160691 are highly
                   unstable. The orbital parameters of 
                   the HD$\thinspace$12661 planets are located in a border region between stable and
                    unstable dynamical regimes, so while its currently determined orbital parameters 
		    produce stable regular orbits, a minor change within the
                   margin of error of just one parameter may result in a chaotic dynamical system.

                   \end{abstract} 


                   \section{Introduction}
                   The recent explosion in the number of newly detected extrasolar planets has
                    brought
                   the total number of such planets to about one hundred; it appears that the 
                   exact number changes too rapidly to be quoted. Some of the planets form 
                   multiple planetary systems around their parental stars. At the moment the 
                   total number of systems with two or more planetary-mass companions around 
                   main sequence stars has reached 11. 
                    Four planetary systems - HD$\thinspace$12661, HD$\thinspace$38529,  HD$\thinspace$37124 
                   and HD$\thinspace$160691 - have recently been elevated 
                   to the status of multiple, following the discovery of a second planetary
                    companion 
                   in each of them. In most of these systems the strong dynamical interaction 
                   between planets makes planetary orbital parameters (see Table 1),
                   found using standard two-body Keplerian fits, unreliable (Laughlin \& Chambers
                   2001, 2002). There is also a great uncertainty in the determination of planetary
                    masses. All 
                   those leave us a substantial available parameter space to be explored in 
                   order to exclude the initial conditions which lead to dynamically unstable 
                   configurations. 

                   A classical method that allows one to distinguish between regular
                   and chaotic dynamical states is the method of Lyapunov Characteristic Numbers
                    (LCN).
                   Let us note that {\it chaotic} in the Poincar{\'e} sense means that the dynamical
                    behavior is
                   not quasi-periodic, and does not 
                   necessarily mean that the system will disintegrate during any limited period of
                    time.
                    The estimation of LCN usually requires computations over long 
                   evolutionary time, sometimes much longer than the lifetime of the system 
                   studied.   

                   In our previous papers (see for example, Go{\'z}dziewski et al. 2001a, 2001b) we
                    showed 
                   that a new method developed by Cincotta \& Sim\'o (2000) and called {\bf MEGNO}
                   (the acronym of {\it Mean Exponential Growth of Nearby Orbits}), 
                   can be successfully applied to the studies of dynamical stability of 
                   extrasolar planetary systems. This method is based on the same ideas as LCN 
                   but converges about 100 times faster, and is more sensitive. For example its application 
                   to the Gliese 876 system (Go{\'z}dziewski et al. 2001b) clearly identified the 
                   exact location of the 2:1 mean motion resonance and its width. 
                   MEGNO helped to show (Go{\'z}dziewski \& Maciejewski 2001) that while the
                    orbital parameters of the HD 82943 system 
                     derived from the Keplerian fit lead to an unstable self-destructing system,
                    there are some small changes 
                     in this fit which lead to stable configurations.

                    In this work we apply the  technique to all four new planetary systems. 
                   We are especially interested in a comparative study of the global dynamics 
                   of these systems, as they are likely to represent  different types of
                    dynamical 
                   behavior. 
                    Taking into account the rather 
                   preliminary nature of all orbital fits, we hope that this paper will provide a
                    useful guide for available 
                   stable orbital parameters (such as { $a$},{  $e$} and {$\omega$}) 
                   for more sophisticated self-consistent fits (Laughlin \& Chambers 2001, 2002;
		   Marcy et al. 2002). It may also provide tighter constraints
		   on the parameter space available to some planets due to
		   unsufficient number of observations  as in the case of the
		   HD$\thinspace$37124c planet where any eccentricity $e_c$
		   between 0.3 and 0.8 fits the observational data within the
		   velocity errors (Butler et al. 2002).
		    
                   \begin{deluxetable}{llllllll}
                   \tablewidth{0pt}
                   \tablecaption{Orbital parameters of new planetary systems.}

                   \tablehead{
                   \colhead{Name}           & $M_p\sin i$($M_J$)      &
                   $M_*$($\Msun$)  & $a$(AU) 
                    & $P$(days)          & $e$  
                     &
                   $\omega$(deg) & 
		   $T_{peri}^*$(JD-2450000)}

                   \startdata
                   {\bf HD$\thinspace$12661b\tablenotemark{a}} & 2.30& 1.07& 0.82& 263.3& 0.35 & 292.6 & 9943.7 \\
                   {\bf HD$\thinspace$12661c\tablenotemark{a}} & 1.56& 1.07& 2.56& 1444.5& 0.20 & 147.0 & 9673.9\\
                   \hline
                   {\bf HD$\thinspace$37124b\tablenotemark{a}} & 0.86&0.91  &0.54& 153.3& 0.10 & 97.0 & 1227\\
                   {\bf HD$\thinspace$37124c\tablenotemark{a}} & 1.01&0.91  &2.95&1942.0& 0.40 & 265.0 & 1928\\
                   \hline
                   {\bf HD$\thinspace$38529b\tablenotemark{a}} & 0.78& 1.39& 0.13& 14.3& 0.28& 90.0 & 10005.8\\
                   {\bf HD$\thinspace$38529c\tablenotemark{a}} & 12.78& 1.39& 3.71& 2207.4& 0.33& 13.0 & 10043.7\\
                   \hline
                   {\bf HD$\thinspace$160691b\tablenotemark{b} } & 1.7$\pm$0.2& 1.08& 1.5$\pm$0.1
                    & 638$\pm$10&0.31$\pm$0.08& 320$\pm$30 & 50698$\pm$30\tablenotemark{c}\\
                   {\bf HD$\thinspace$160691c\tablenotemark{b}} & 1.0 &1.08& 2.3& 1300& 0.8& 99 & 51613\\

                   \enddata
 		   \tablenotetext{*}{In our calculations we use as an initial orbital parameter the Mean Anomaly 
which is a function of $T_{peri}$}
                   \tablenotetext{a}{Data from  http://www.exoplanets.org as on August 10,
		   2002}
                   \tablenotetext{b}{Data from Jones et al. 2002}
		   \tablenotetext{c}{HJD}

                   \end{deluxetable}


                   \section{Results.}

                   All our results obtained with MEGNO were confirmed 
                   by direct integration of the equations of motion using the Lie-series method
                   (e.g. Hanslmeier \& Dvorak, 1984) which has been already
                   used for many numerical simulations of Solar System dynamics (Tsiganis et al.
                    2001) and
                   dynamics of extrasolar planetary systems (Pilat-Lohinger \& Dvorak 2002). 
                   This integration method is based on
                   recurrence formulae and uses an automatic step-size control. 
                   The four systems under  study
                   were integrated over evolutionary times of up to $10^7$ yrs. In all our
		   models we considered the Keplerian orbital elements from Table 1 in
		   reference to the 
		   coordinate system centered at the central star. However we are aware that 
		   for multiple planet systems orbital elements derived from summed 2-body 
		   fits to a radial velocity data might be better interpreted as referring 
		   to Jacobi coordinates (see, for example, Rivera \& Lissauer 2000).
                   We will try this approach in our future studies.
		   Our study 
                   revealed many interesting and rather unexpected dynamical features in all of
                    these systems.
                   It is remarkable that being discovered almost simultaneously they represent a
                    real
                   `dynamical zoo' that we describe below. Let us note that because of
		   the preliminary nature of the orbital parameters in all systems we discuss
		   in this paper, conclusions about  the exact dynamical status of a particular system may
		   differ from results presented here as the determination of orbital parameters will
		   continue to be improved. However, we expect that the method we use and the general
		   conclusions about dynamical properties of the parameter space available for each system will
		   have a longer term value.

                   \subsection{HD$\thinspace$12661}

                   In  HD$\thinspace$12661 the two planets  are rather close to each other. In fact, for this
                    system 
                   the ratio of the periastron distance 
                   of the outer orbit to the apastron distance of the inner orbit is very close to
                   the corresponding critical value for three-body dynamical stability derived by
                    Eggleton \& Kiseleva (1995)
                   \footnote{Note that in their Eq2 the second sign should be "-"}
                   for less extreme mass ratios (triple stars). This places HD$\thinspace$12661 in a very
                    interesting zone 
                   of critical parameters
                    where small changes can lead to totally different dynamical regimes.
                   Fig.1 shows the variations of the semi-major axis {$a_b$} of the innermost planet
                    in HD$\thinspace$12661
                   over 1 million years. In the lower panel, which represents the system with nominal
                    initial orbital
                   parameters (Table 1), $a_b$ displays a regular quasi-periodic behavior with no
                    sign of 
                   unpredictable variations over time.  The top panel shows totally different behavior of { 
                    $a_b$} for 
                   the same planetary system with  only a very small difference in initial value of
                    { $a_c$}: 
                   $a_c$=2.58 {AU} instead of 2.56 {AU}.  This $\Delta a_c = 0.02$ {$\thinspace$AU} 
                   is well within the error of the { $a_c$} determinaton (see, for example, Fischer
                    et al. 2001). In this `new' system the semi-major axis $a_b$ displays 
                   irregular and therefore 
                   unpredictable jumps from one `mean' value to another. We call such an
                    orbit chaotic. 
                   This phenomenon of unpredictibilty
                   is well-known in dynamical systems since Poincar{\'e}, and it is a characteristic 
                   of chaos in such systems. 
                   These results are in a very good
                    agreement with MEGNO tests: for the system with the nominal parameters of 
                   HD$\thinspace$12661 $a_c$=2.56 {$\thinspace$AU}
                   the MEGNO indicator (see Go{\'z}dziewski et al. 2001a,b)
                    converges 
                   to the value 2 characteristic for regular systems,  
                   while for the system with $a_c$=2.58$\thinspace$AU
                   $<$Y$>$ is $\sim$6, 
                   which indicates the presence of chaos in this system. Let us note we do not
                    claim that 
                   such a system will disintegrate after any particular period of time. We just
                    point out that 
                   its dynamical state over a long time is unpredictable. From the conventional point
		   of view one would expect the system with larger $a_c$ to be more stable. However, this
		   conventional wisdom fails for  systems whose orbital parameters are located within 
		    regions of marginal stability in the phase space (see below). 

                   Fig.2 shows the MEGNO stability map for initial $a_b$-$a_c$ of
                    HD$\thinspace$12661  in a rather
                   large neighbourhood of the nominal values of these orbital parameters. Filled circles mark stable 
                   regular systems  where $<$Y$>\in$[1.97.203]. Small dots mark chaotic regions where $<$Y$> <$ 1.95
			or $<$Y$> >$ 2.05. Open circles with dots mark intermediate cases. One can see
                    that the stability  map itself
                   looks rather chaotic, as it displays a number of  
		   chaotic initial orbital parameters (marked with open circles and with small dots) 
		   unsystematically located within fairly large regular zones 
                   (), marked with the filled circles.  
                    This
                    particular 
                   structure of the phase space is  typical for regions close to the border 
                   between stable and unstable motions (e.g. Dvorak et al. 1989). In such regions 
                   minor changes in each of the parameters may dramatically transfer the system from
                    a regular into a chaotic dynamical state.
                   This behavior is not
                   observed in the motion of the planets in our Solar System  (e.g. Murray \&
                    Holman 2001), 
                   but it is a well-known phenomenon for the dynamics of asteroids (Wisdom 1983). 
                   Kiseleva et al. (1994a, 1994b) found a similar phenomenon for simulated triple
                    stars
                   near the stability limit. The 9:2 mean motion resonance may 
		   be responsable for the rather wide strip of chaotic parameters.
		    
                   \begin{figure}
                   \centering
                   \includegraphics[totalheight=17cm]{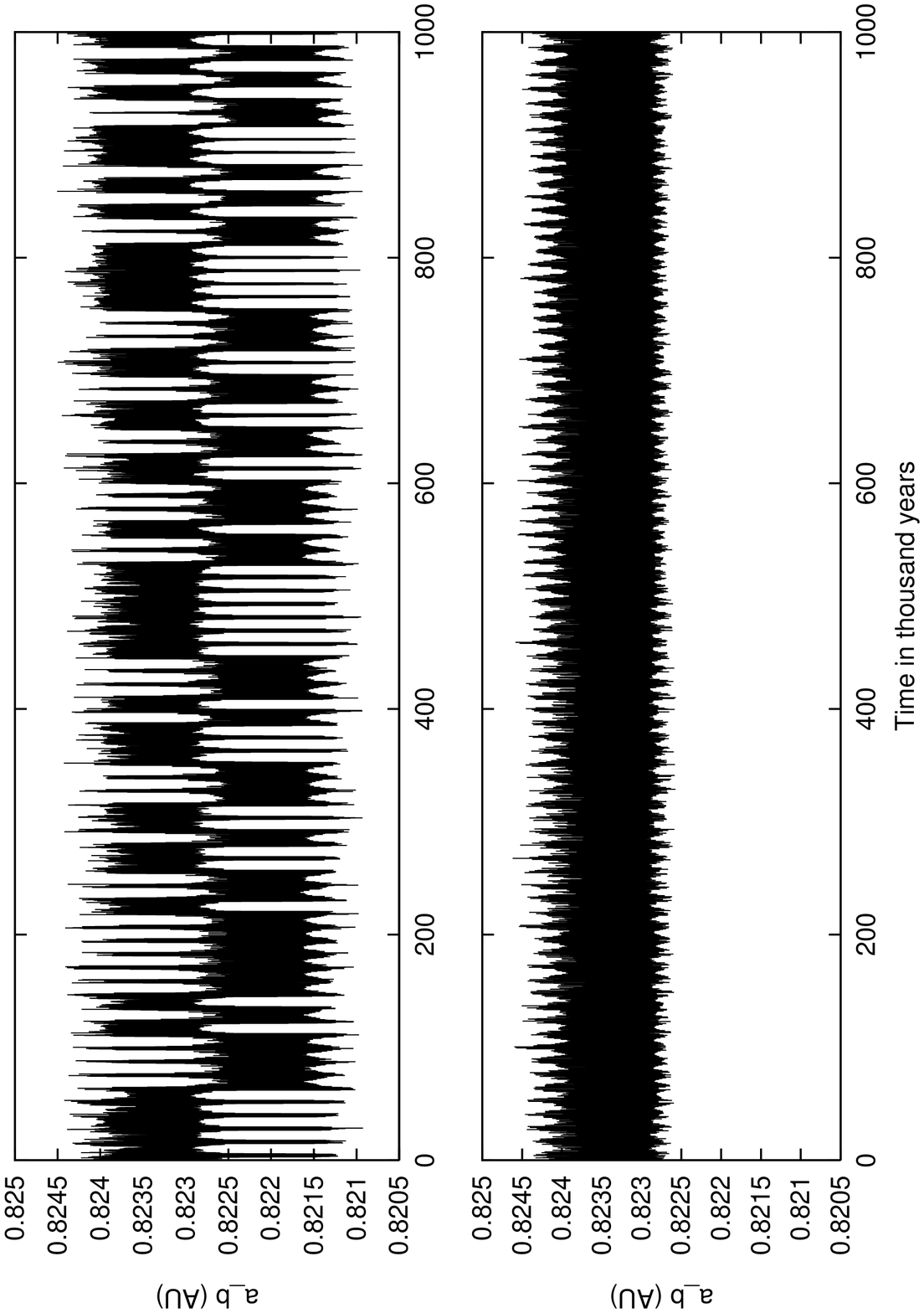}
                   \caption{Variations of semi-major axis $a_b$ of the innermost planet in HD$\thinspace$12661
                    over 
                   one million years: in the top panel the semi-major axis $a_c$ of the outer
                    planet 
                   is {\bf 2.58} {AU} and in the lower panel $a_c$ = {\bf 2.56} {$\thinspace$AU}. All other
                    parameters are the 
                   same in both panels.} 
                   \end{figure}  
                     
                   \begin{figure}
                   \centering
                   \includegraphics[totalheight=17cm, width=15cm]{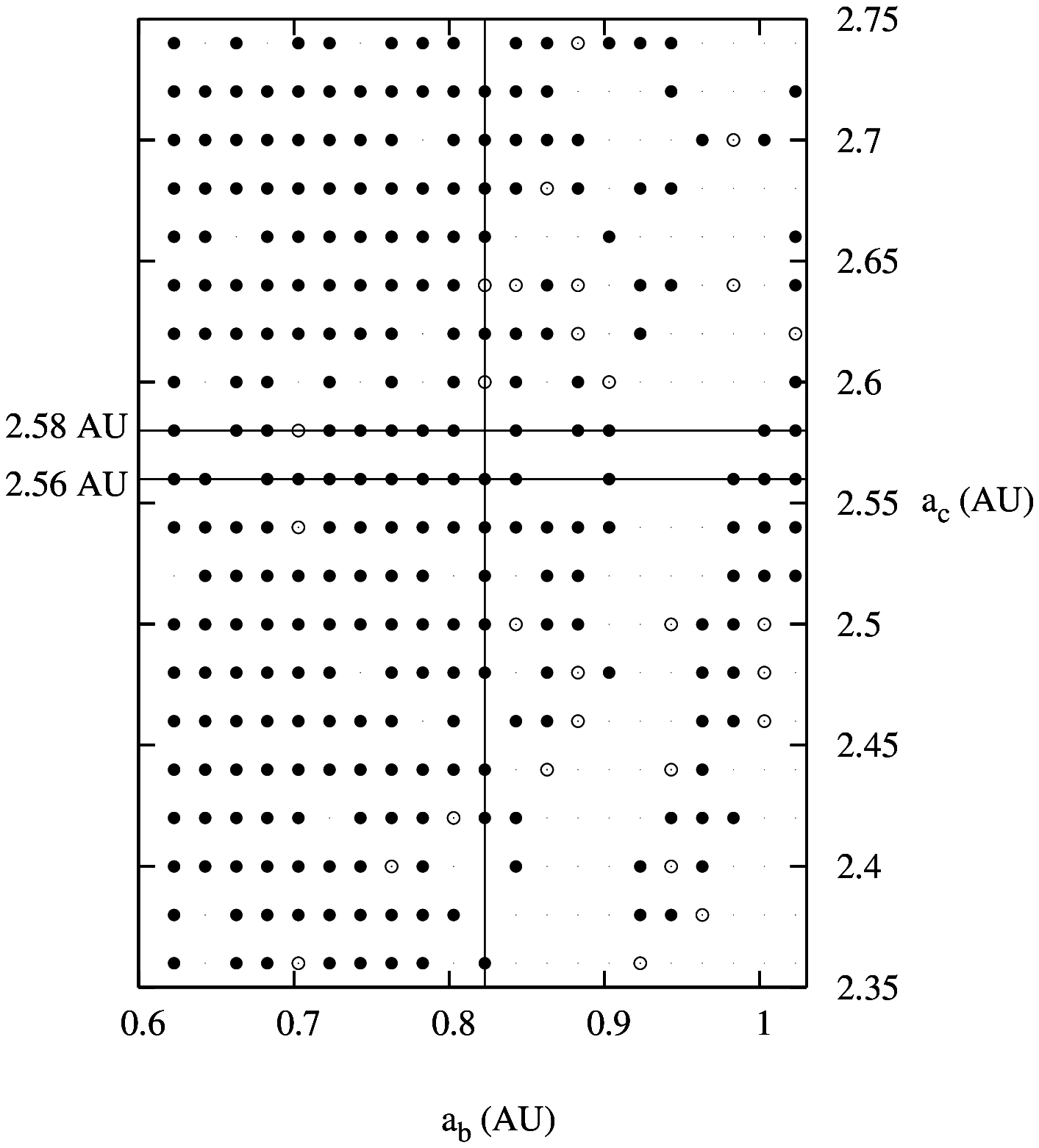}
                   \caption{ Initial semi-major axes leading to regular orbits (filled circles) and chaotic 
                   orbits (open circles and small dots) in the HD$\thinspace$12661 planetary system. All
                    other orbital initial parameters 
                   are equal to the nominal parameters of HD$\thinspace$12661 (Table 1).
                   The orbital parameters with $a_c$=2.56 and 2.58$\thinspace$AU discussed in the text are 
		   indicated by
		   line intersections. 
                    The wide chaotic strip corresponds to the 9:2 mean motion resonance.}
           
                   \end{figure}

                   \subsection{HD$\thinspace$37124 and HD$\thinspace$38529}

                   The orbits in both these systems are solidly stable with no
                    chaotic features for
                   a wide range of orbital parameters including high relative inclinations between
                    the two planetary 
                   orbits. Let us note that all results presented in this paper were obtained for
                    nearly coplanar systems
                   unless it is indicated overwise. When applied to the planetary orbits of the HD
                    38529 system
                    with different variations of its orbital parameters, the MEGNO indicator
                    $<$Y$>$ almost always converges to 2,
                    with a few exceptions for very large eccentricities $e_c$ of the outer planet
                    (see Fig. 3c). 
                   Direct N-body integrations of the equations of motion performed for selected
                    samples also
                   showed a quasi-periodic behaviour for all orbital parameters considered. The
                    same is true for HD$\thinspace$37124: 
                   this system becomes unstable when $e_c$ approaches 0.6 (see Fig. 3b). Butler et al. (2002)
		   pointed out that with only 30 observations over more than 5 years the eccentricity 
		   of the outer planet $e_c$ in this system is poorly 
		   constrained and can be anything between 0.3 and 0.8. Our results are in very good 
		   agreement (Fig. 3c) with their critical for stability value $e_c\sim 0.65$ for co-planar 
		   configurations with $\sin(i)$=1.0. This is precisely the case when MEGNO can be 
		   used much more efficiently than standard numerical experiments to set 
		    tight dynamical constraints on the uncertain parameters of planets.	   
                   
                   However, the orbital parameters of planets in both systems, 
                   especially the planetary eccentricities (which are the orbital
                    parameter 
                   most sensitive to external perturbations), are not  unaffected by the
                    presence
                   of the other planetary companion. For example, the eccentricity of the planet HD
                    37124b 
                   changes periodically between 0.1 and 0.43. Even in the HD$\thinspace$38529 system where the planets are 			not expected
                    to have much influence on
                   each other's dynamics, the eccentricity of the very close to the star
                    inner planet ($P_b=14.3$ days) fluctuates within the interval $[0.292,
                    0.368]$ for 
                   a nearly coplanar model with nominal parameters. The massive outer planet
		   with nearly substellar mass $M_c^{min}=12.78 M_J$       
                   is also subjected to small eccentricity fluctuations. 
                    
                   It can be interesting to consider a case where the two planetary planes in HD
                    38529 have  high relative 
                   inclination. MEGNO tests showed that the orbits in this system remains stable
                   for {\it all} relative inclinations, including orthogonal orbits with
                    $i_r=90^{\rm o}$. Taking into account 
                   the sub-stellar minimal mass of planet C, completely independent formation
                    scenarios for 
                   companions B and C are possible. Therefore 
                   a highly non-coplanar configuration would not be unlikely for this system.
                    The innermost planet B is close enough to the star for tidal 
                   friction to become important when the `Kozai effect' (Kozai 1962)  induces 
                   large fluctuations on the orbital eccentricity ($e_{max}=$1 for $i_r=90^{\rm o}$). 
                   This may significantly affect the dynamics of this system 
                   (see, for example, Kiseleva et al. 1998; Eggleton et al. 1998; 
                   Eggleton \& Kiseleva-Eggleton 2001a, 2001b; Wu \& Murray, in preparation). We
                    are going to investigate 
                   these effects in a future paper.

                   \subsection{HD 160991}

                   Our simulations show that for HD$\thinspace$160691 
                   the initial orbital parameters in Table 1 lead to a highly
                    unstable self-destructing system. 
                   However, the orbital parameters of both planetary orbits are rather speculative
                    at the moment: even 
                   the existence of the second planetary companion is not yet fully confirmed
                    (Jones et al. 2002). 
                   Therefore we looked for stable configurations among  wide ranges of orbital
                    parameters. We found a few stable configurations  located not very far in the
		    parameter space from the currently defined parameters of the system (see Fig. 3d).  
                   It looks as if a high eccentricity of the outermost planet $e_c>0.7$ is an important
		   stabilizing factor in this system.  
                   Using MEGNO we were able to identify a few stability zones in a parameter
                    space which includes together 
                   with the parameters listed above also parameters not determined from
                    observations such as 
                   the relative inclination $i_r$ between the two planetary orbits. 
                   Most of these stable configurations are associated with the
		   2:1 mean motion resonance. We are going to discuss in detail the possible combinations of
                    parameters resulting in stable 
                   orbits in  HD 160991 in our forthcoming paper (Bois et al., in preparation).

                   \section{Comparative stability analysis and conclusions}

                   \begin{figure*}
                   \centering
                   \includegraphics[totalheight=19cm, width=15cm]{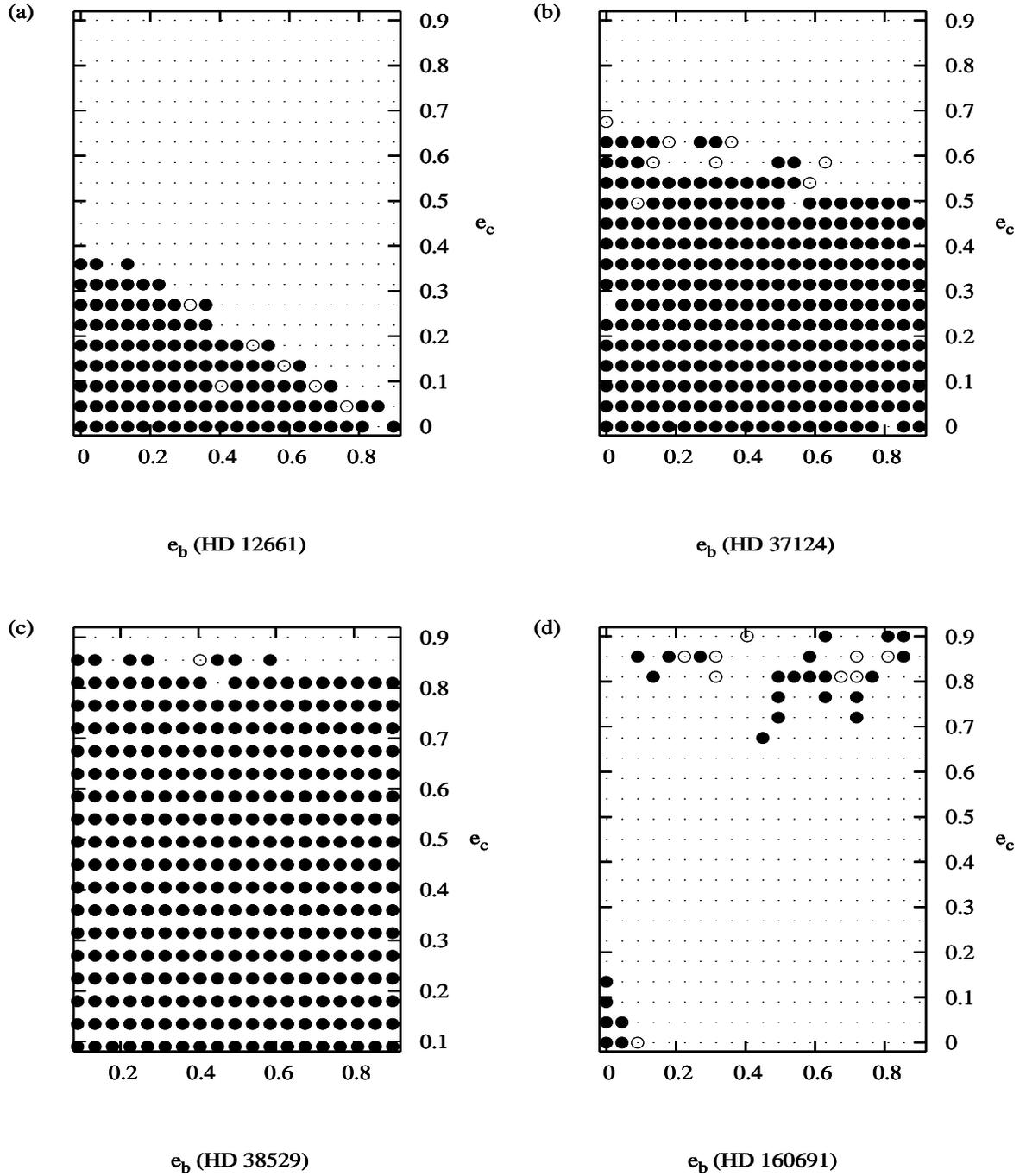}
                   \caption{Stability maps in the $e_b - e_c$ parameter space for all
                    four new planetary systems. 
                   The symbols  are the same as in Fig. 2 (filled circles indicate stable
                    regular orbits).}
                   \end{figure*} 

                   Fig. 3 presents a comparative visualisation of our MEGNO stability analysis for
                    all four new planetary systems.
                   For this visual presentation we choose to consider orbital stability as a
                    function of both orbital eccentricities
                   simply because the values of $e_b$ and $e_c$ can only be changed between 0
                    and 1, and 
                   therefore do not require any scaling (as for example would semi-major axes) for
                    different systems. All other
                   orbital parameters are the nominal ones from Table 1. It is easy to see 
                   the differences in dynamical status of the four systems and to identify the
                    ranges of eccentricities which 
                   allow stable planetary orbits in each of them. We also produced simular maps 
		   for other pairs of orbital 
                   parameters. Such ranges of stable parameters can be very
                    useful 
                   for the improvement of preliminary orbits in the new planetary  systems
                    discussed in this paper. In our future papers we plan to present 
                   detailed dynamical analyses of each system taking into account 
                   angular orbital parameters not constrained by observational data ($i_r$x,
                    $\Omega$), as well as
                   $\sin i$ and the resulting different planetary masses. 

                   \acknowledgments

                   This work was supported by the John Templeton Foundation (Agreement 
		   ID \#398-COS272). 
		   RD thanks the Austrian FWF P14375-TPH for a support in travel 
		   expenses.
		   LK-E thanks the University Bordeaux 1 for a research fellowship.
                   LK-E and RD thank the Bordeaux Observatory for its hospitality. All authors thank 
		   the referee Greg Laughlin for very useful comments and suggestions.

                   \end{document}